\documentclass{PoS}

\usepackage{subfig}

\def\Nf{N_\mathrm{f}}
\def\gA{g_\mathrm{A}}
\def\ZA{Z_\mathrm{A}}
\def\gT{g_\mathrm{T}}
\def\mN{m_\mathrm{N}}
\def\mps{m_\mathrm{PS}}
\def\fps{f_\mathrm{PS}}
\def\csw{c_\mathrm{SW}}
\def\ksc{\kappa^{(S)}_c}
\def\msbar{\overline{\mathrm{MS}}}

\title{Nucleon form factors and structure functions from
      $\Nf=2$ clover fermions}

\ShortTitle{Nucleon form factors and structure functions}

\author{
  QCDSF/UKQCD Collaboration:
  S.~Collins,$^a$
  M.~G\"{o}ckeler,$^a$
  Ph.~H\"{a}gler,$^a$
  T.~Hemmert,$^a$
  R.~Horsley,$^b$
  Y.~Nakamura,$^{a,c}$
  A.~Nobile,$^a$
  H.~Perlt,$^d$
  \speaker{D.~Pleiter},$^e$
  P.E.L.~Rakow,$^f$
  A.~Sch\"{a}fer,$^a$
  G.~Schierholz,$^e$
  A.~Sternbeck,$^a$
  H.~St\"{u}ben,$^g$
  F.~Winter$^a$ and
  J.M.~Zanotti$^b$\\
  \llap{$^a$} Institut f\"ur Theoretische Physik, Universit\"at Regensburg,
              93040 Regensburg, Germany \\
  \llap{$^b$} School of Physics, University of Edinburgh, Edinburgh EH9 3JZ,
              UK\\
  \llap{$^c$} Center for Computational Sciences, University of Tsukuba,
              Ibaraki 305-8577, Japan \\
  \llap{$^d$} Institut f\"ur Theoretische Physik, Universit\"at Leipzig,
              04109 Leipzig, Germany \\
  \llap{$^e$} Deutsches Elektronen-Synchrotron DESY, 15738 Zeuthen, Germany \\
  \llap{$^f$} Theoretical Physics Division, Department of Mathematical Sciences,
              University of Liverpool, Liverpool L69 3BX, UK \\
  \llap{$^g$} Konrad-Zuse-Zentrum f\"ur Informationstechnik Berlin,
              14195 Berlin, Germany \\
  Email: \email{dirk.pleiter@desy.de}
}

\abstract{
We give an update on our ongoing efforts to compute the nucleon's
form factors and moments of structure functions using
$\Nf=2$ flavours of non-perturbatively improved Clover fermions.
We focus on new results obtained on gauge configurations where the
pseudo-scalar meson mass is in the range of 170-270~MeV.
We will compare our results with various estimates obtained from
chiral effective theories since we have some overlap with the quark
mass region where results from such theories are believed to be applicable.
}

\FullConference{%
  The XXVIII International Symposium on Lattice Field Theory, Lattice2010\\
  June 14-19, 2010\\
  Villasimius, Italy
}

\begin{document}

\section{Introduction}

Over years significant efforts have been made to use lattice techniques to
investigate the structure of the nucleon. Of particular interest are the
Parton Distribution Functions (PDFs) and form factors. The latter encode
information about charge distribution and magnetization while the
PDFs tell us about the distribution of momentum and spin.
While some of the related observables can be determined with good accuracy
by experiments (e.g.~the nucleon's axial charge $\gA$) other quantities are
difficult to access (like the tensor charge $\gT$).

A precise determination of moments of nucleon PDFs and form factors on
the lattice turned out
to be rather challenging. It continues to be difficult to reach sufficient
control on all systematic errors such as finite size effects, lattice artefacts 
and the influence of the chiral extrapolation.
Simulations are performed in volumes
of a size where some quantities exhibit significant finite size effects.
The available lattice data of the quantities of interest show no significant
discretization effects. But current simulations only probe a small
window of lattice spacings thus providing us with limited control on the
continuum extrapolation. From chiral effective theories (ChEFT) there are
indications that the quark mass dependence close to the physical pion mass
is very strong. Therefore extending lattice simulations into the region
where $m_\pi \le \mps \lesssim 300~\mbox{MeV}$ has become a major goal
for recent calculations.

\section{Simulation details}

For our simulations we use Wilson glue and $\Nf=2$ degenerate flavours of
Clover fermions, where the improvement coefficient $\csw$ has been determined
non-perturbatively. Most of our configurations have been generated using
the BQCD implementation of the HMC algorithm \cite{bqcd}. Various
algorithmic improvements have been applied which accelerate this
algorithm, such as the Hasenbusch preconditioning and the use of multiple
time-scales, or reducing the time spent for matrix inversion. For
instance, chronological guess and the Schwarz Alternating Procedure
(SAP) are used to start the inversion and to precondition the fermion
matrix \cite{andrea}.

\begin{figure}[t]
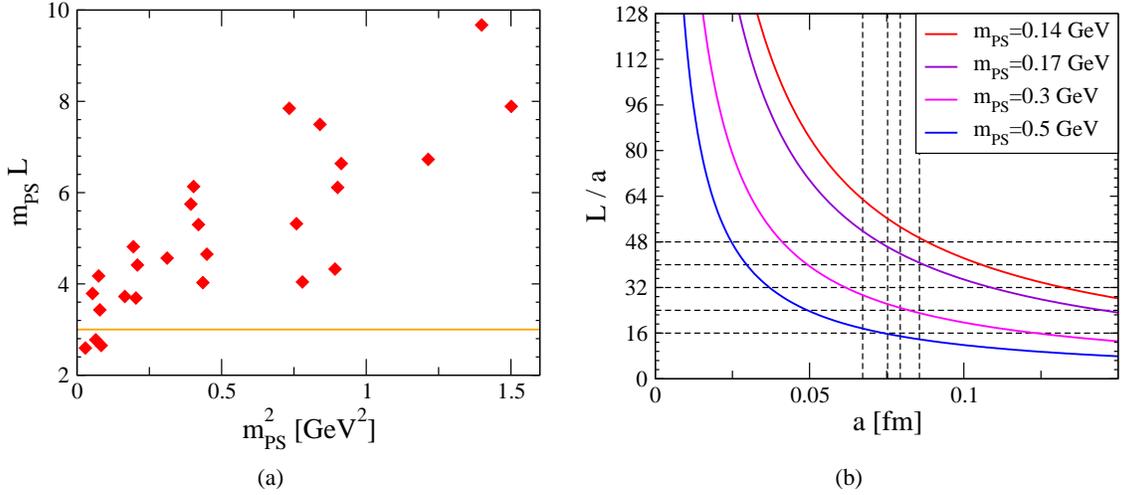

\vspace*{-5mm}
\subfloat[]{
  \label{fig:mpsL-vs-mps2}
  \includegraphics[scale=0.5]{figs/mpsL-vs-mps2}
}
\hfill
\subfloat[]{
  \label{fig:ns-vs-a_mpsL=3}
  \raisebox{2mm}{\includegraphics[scale=0.5]{figs/ns-vs-a_mpsL=3}}
}
\vspace*{-2mm}
\caption{The left panel shows the simulation points in the $\mps^2$
vs.~$\mps L$ plane. In the right panel dashed lines show the lattice spacing
and box sizes for which simulations have been performed. In both figures
the continuous lines show where $\mps L=3$.
}
\end{figure}

These algorithmic improvements plus recent increase in computing resources
enabled simulations in the region of small quark masses, i.e.~in a region
where the pseudo-scalar mass is smaller than 300~MeV.
Fig.~\ref{fig:mpsL-vs-mps2} shows the parameter region of our simulations.
When approaching physical quark masses larger lattices are needed to stay
in the region $\mps L\gtrsim 3$ where finite size effects are expected to
be sufficiently small (see Fig.~\ref{fig:ns-vs-a_mpsL=3}). To investigate
such finite size effects we have also performed simulations with $\mps L < 3$.

We compute the quark propagators using point sources which we (Jacobi)
smear to improve overlap with the ground state.  For the three-point
correlation functions we apply standard sequential source techniques.
The distance between source and sink is about
1~fm.
Throughout this paper we will ignore contributions coming from disconnected
terms. While these anyhow cancel in the iso-vector cases, results for the
iso-scalar case maybe affected by an uncontrolled systematic error.

To set the scale we use the Sommer parameter $r_0/a$ which we extrapolated
to the chiral limit at each beta.
While on the lattice this quantity can be determined
with small statistical errors, there is no experimental determination.
We therefore computed the dimensionless quantity $(a m_N) (r_0/a)$ on the
lattice and use the experimentally well known mass of the nucleon $m_N$ to
obtain $r_0 = 0.467\,\mbox{fm}$.

Most of the quantities considered in this paper need to be renormalised.
The renormalisation constants have been determined using the
$\mathrm{RI}^\prime\mathrm{-MOM}$ scheme
\cite{Z}, except for the vector current renormalisation constant $Z_V$.
Here we applied the condition that the nucleon's local vector current at
zero momentum must be 1. If necessary, the results are converted into
$\msbar$ scheme using the 4- and 2-3-loop expressions of the
$\beta$ function and corresponding anomalous dimension $\gamma$, respectively.

\section{Lowest moments of PDFs}

Let us first consider the lowest moment of the polarized nucleon PDF
$\langle 1\rangle_{\Delta q}$ (also known as axial coupling constant $\gA$).
This quantity is determined from the renormalised axial
vector current $A_{\mu}^R = Z_A\,(1+b_A\,a m_q) A_{\mu}$, where $a m_q =
(1/\kappa - 1/\ksc)/2$. $Z_A$ is known non-perturbatively \cite{Z},
for $b_A$ we use a tadpole improved one-loop perturbation theory result.

\begin{figure}[t]
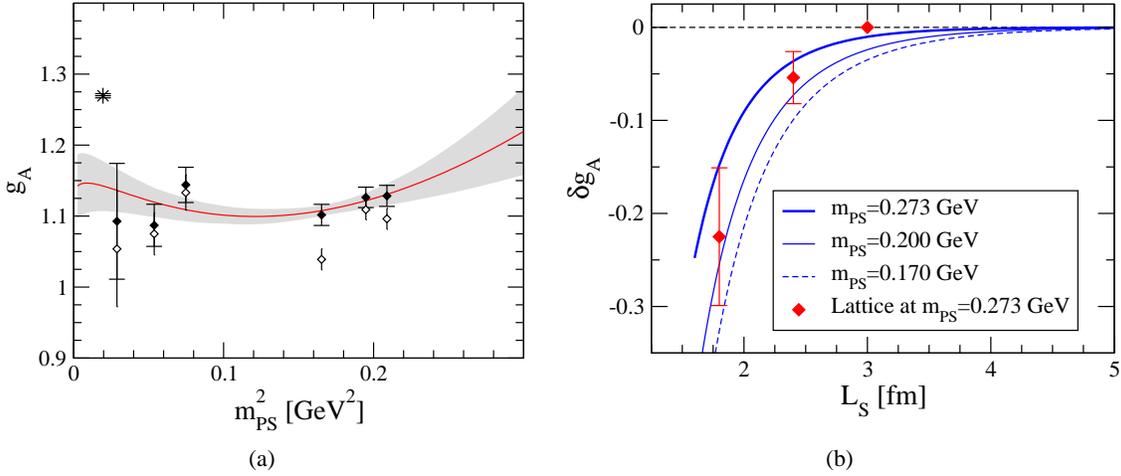

\vspace*{-5mm}
\subfloat[]{
  \label{fig:fitGaBchpt}
  \includegraphics[scale=0.5]{figs/fitGaBchpt}
}
\hfill
\subfloat[]{
  \label{fig:gA-fse}
  \raisebox{2mm}{\includegraphics[scale=0.5]{figs/gA-fse}}
}
\vspace*{-3mm}
\caption{The left panel shows $\gA$ as a function of $\mps^2$. The
open and filled diamonds show the lattice results before and after correction
of finite size effects, respectively.
The star indicates the experimental result.
The line shows a fit to the data as described in the text.
The right panel shows the relative finite size effects determined on the
lattice (symbols) and obtained from a fit to an expression from ChEFT.
}
\end{figure}

We have fitted our lattice results
to an expression from ChEFT based on the
SSE formalism. Using this formalism both the quark mass dependence
\cite{qcdsf-ga-mq} and the finite volume dependence \cite{qcdsf-ga} have
been calculated. Since our results for different lattice spacings do not
exhibit clear discretization effects we combine all our results where
$\mps \le 450\,\mbox{MeV}$. The fit range has been chosen such that stable
fits are obtained. Our data is not sufficiently precise to determine all
parameters. We therefore fix a few parameters to their phenomenological value
and keep only $\gA$ in the chiral limit, the leading $\Delta\Delta$-coupling
$g_1$ and the SSE coupling term $B_9^r(\lambda)$ as free fit parameters.
The resulting fit and the lattice data are shown in Fig.~\ref{fig:fitGaBchpt}.

In our fit we only included results for the largest lattice at a given
set of bare parameters. For some data sets we have results
for different volumes. We thus can compute the relative shift
\begin{equation}
\delta_{\gA}(L) = \frac{\gA(L) - \gA(\infty)}{\gA(\infty)}
\end{equation}
both from the fit as well as from the lattice data
taking the results on the largest lattice as approximation of $\gA(\infty)$.
In Fig.~\ref{fig:gA-fse} we compare the relative shift for different values
of the quark mass with our lattice results at $\mps \simeq 270\,\mbox{MeV}$.
The shift predicted from ChEFT only slightly underestimates the relative
shift computed on the lattice.

\begin{figure}[ht]
\subfloat[]{
  \label{fig:gA_fps}
  \includegraphics[scale=0.5]{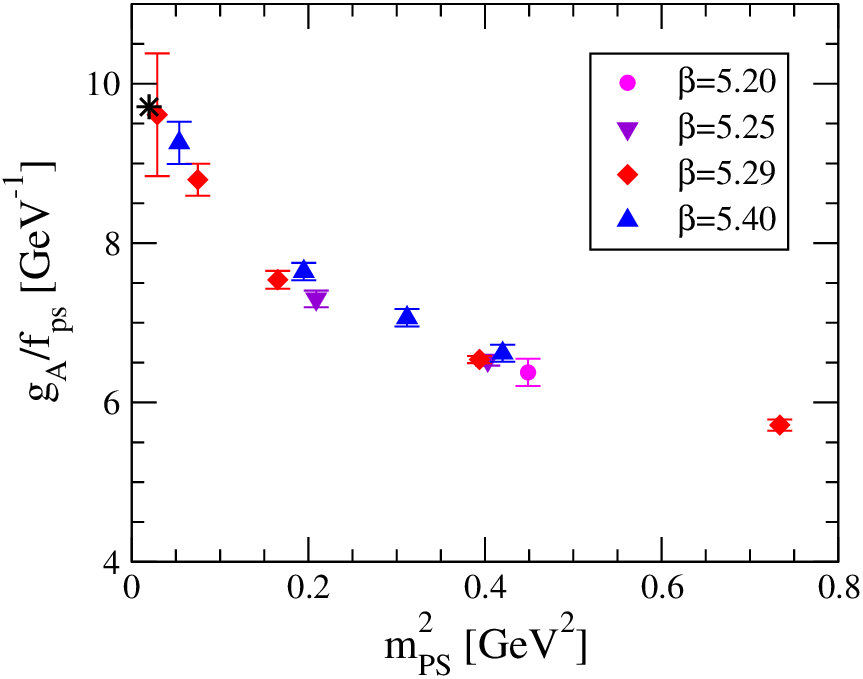}
}
\hfill
\subfloat[]{
  \label{fig:gT_raw}
  \raisebox{0mm}{\includegraphics[scale=0.49]{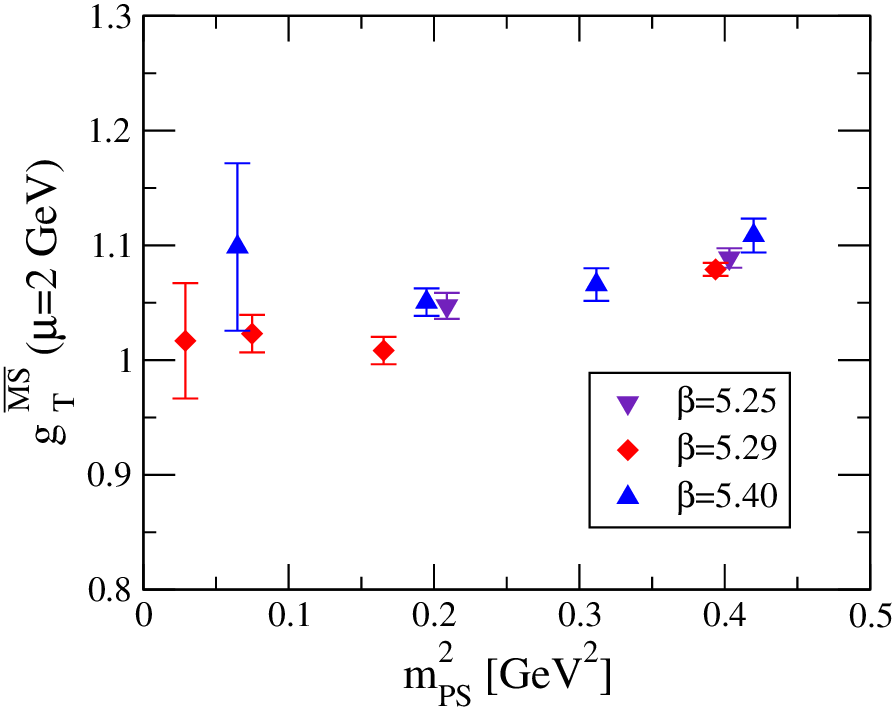}}
}
\vspace*{-3mm}
\caption{The left panel shows $\gA/\fps$ as a function of $\mps^2$.
The right panel shows our results for $\gT^{\msbar}$ at a scale
$\mu=2\,\mbox{GeV}$.
}
\end{figure}

Also after correcting for finite size effects we observe a significant
difference to the experimental value. It is interesting to notice that
a much better agreement with the experimental value is observed for the
ratio $\gA / \fps$ (see Fig.~\ref{fig:gA_fps}).
In this ratio the renormalization constant $\ZA$ drops out.

In Fig.~\ref{fig:gT_raw} we show our results for the nucleon tensor charge
$\langle 1\rangle_{\delta q} = \gT$.
We observe only a very mild quark mass dependence and the data
reveals no systematic discretization effects.
This quantity is not well known experimentally.
Our values are larger than the phenomenological results presented
in \cite{Anselmino:2008jk}.

\section{$n=2$ moments of PDFs}

\begin{figure}[hb]
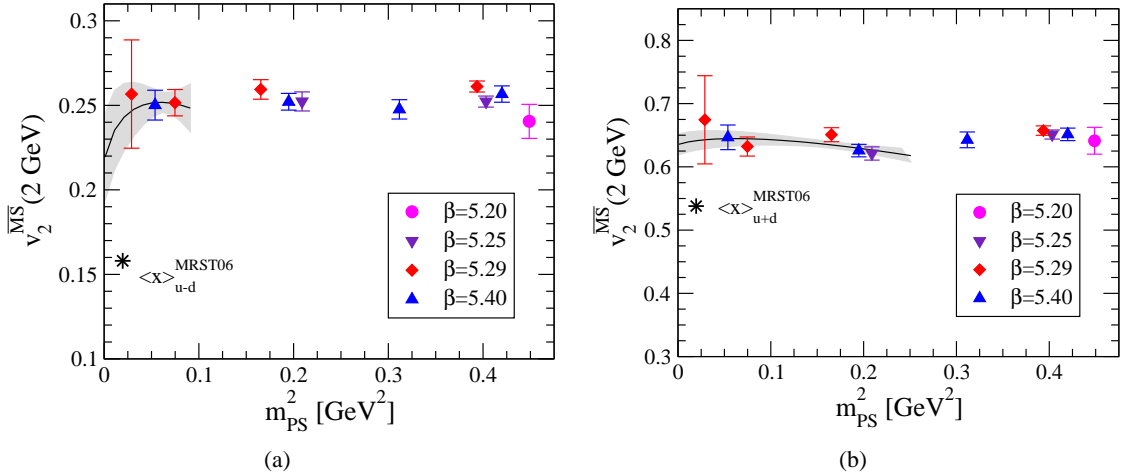

\subfloat[]{
  \label{fig:v2b}
  \includegraphics[scale=0.5]{figs/v2b}
}
\hfill
\subfloat[]{
  \label{fig:v2bs}
  \raisebox{0.5mm}{\includegraphics[scale=0.49]{figs/v2bs}}
}
\vspace*{-3mm}
\caption{The left and right panel show results for the second moment of the
iso-vector and iso-scalar unpolarized PDFs, respectively,
as a function of $\mps^2$.
The solid lines show the fits to an expression from ChEFT.
}
\end{figure}

The lowest moment of the unpolarized PDF $\langle x\rangle_q = v_2$
corresponds to the momentum fraction carried by the quarks in the nucleon.
Lattice results from different collaborations tend to be
significantly larger than the phenomenological value.
Fig.~\ref{fig:v2b} and \ref{fig:v2bs} show our most recent results
for the iso-vector and iso-scalar channel. In the latter case disconnected
contributions have been ignored.

Also shown are the results from a fit to results
utilizing methods of covariant Baryon Chiral Perturbation Theory (BChPT)
\cite{Dorati:2007bk}. Fits have been performed with most parameters
fixed to phenomenologically known values.
The iso-vector (iso-scalar) channel data is fitted with only 2 free
parameters: $v_2$ in the chiral limit and the coupling $c_8$ ($c_9$).
Near the physical light quark masses, BChPT predicts $v_2$ to become
larger when the quark mass becomes heavier. In our data for
$\mps \lesssim 250\,\mbox{MeV}$ we do not see any indication for a
bending down when approaching the physical pion mass.
It thus does not seem that a lack of results at sufficiently small quark masses
could explain the large discrepancy between the phenomenological
value and the lattice results.
There are some indications that part of the discrepancy can be explained
by excited state contamination \cite{x-future}.

In Fig.~\ref{fig:a1_raw} the results for the second moment of the
polarized PDF $\langle x\rangle_{\Delta q} = a_1$ is shown.
Discretization effects again seem to be absent in data. From a comparison
of the results for different volumes it seems that also finite size
effects are small. Results from Heavy Baryon Chiral Perturbation Theory
(HBChPT) \cite{a1-hbchpt} lead to the following expression:
\begin{equation}
a_1^{(u-d)}(\mps) =
C \left[ 1 - \frac{4 \gA^2 + 1}{2 (4 \pi \fps)^2}
         \mps^2 \ln\left(\frac{\mps^2}{\mu^2}\right)\right] + \cdots
\end{equation}
In Fig.~\ref{fig:a1_raw} we plot this expression using $\mu = \mN$
and $C$ chosen such that it matches the phenomenological value.
The bending down which we observe in our data for
$\mps \lesssim 0.5\,\mbox{MeV}$ is much less than one would expect from 
this HBChPT result.

\section{Electromagnetic form factors}

\begin{figure}[t]
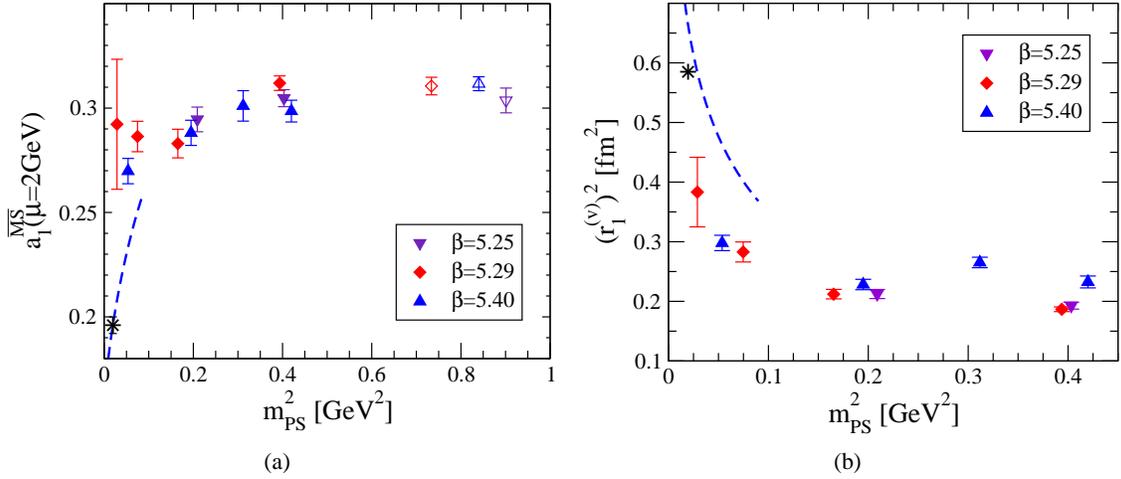

\vspace*{-5mm}
\subfloat[]{
  \label{fig:a1_raw}
  \raisebox{0.4mm}{\includegraphics[scale=0.5]{figs/a1_raw}}
}
\hfill
\subfloat[]{
  \label{fig:r1v}
  \raisebox{0mm}{\includegraphics[scale=0.49]{figs/r1v}}
}
\vspace*{-3mm}
\caption{The left panel shows the results for the second of the
polarized PDFs as a function of $\mps^2$.
In the right panel the results for the Dirac form factor radius $r_1$ are
plotted.  The dashed lines show results from ChEFT as described in the text.
}
\end{figure}

To compute the electromagnetic form factors one makes use of the
standard decomposition of the nucleon electromagnetic matrix elements
%
$\langle p',s' | V_\mu | p,s \rangle =
\overline{u}
\left[
  \gamma_\mu F_1(Q^2) + \frac{\sigma_{\mu\nu}\,q_\nu}{2 \mN} F_2(Q^2)
\right]
u
$,
%
(in Euclidian space) where we use the local vector current $V_\mu$.
$p$ ($s$) and $p'$ ($s'$) denote initial and final momenta
(spins), $q = p' - p$ the momentum transfer (with $Q^2=-q^2$).
To calculate form factor radii and the anomalous magnetic we have to
parametrize the lattice results. Here we use the ansatz
\begin{equation}
F_i(Q^2) = \frac{F_i(0)}{\left[1 + \frac{Q^2}{p m_i^2}\right]^p}
\label{eq:ppole}
\end{equation}
with $p=2$ and $p=3$ for the Dirac and Pauli form factors $F_1$ and
$F_2$, respectively.
Our data is not sufficiently precise to favour a particular parametrization
(see \cite{wolfram-ff} for another parametrization).

\begin{figure}[t]
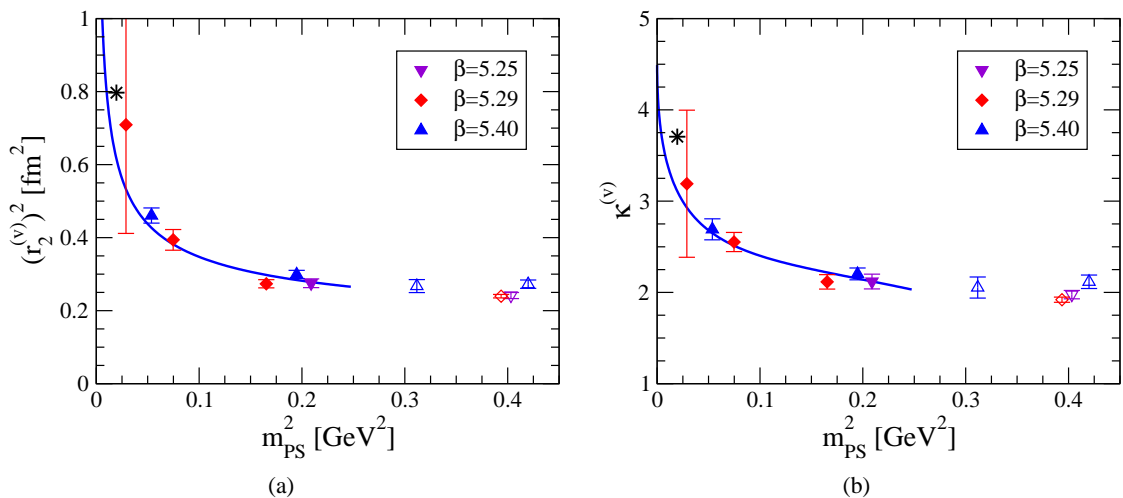

\vspace*{-5mm}
\subfloat[]{
  \label{fig:r2v}
  \includegraphics[scale=0.5]{figs/r2v}
}
\hfill
\subfloat[]{
  \label{fig:kv}
  \raisebox{0mm}{\includegraphics[scale=0.5]{figs/kv}}
}
\vspace*{-3mm}
\caption{The left panel and right panel shows the results for the
Pauli radius $r_2$ and the anomalous magnetic moment $\kappa$.
The solid lines show fits to an expression from ChEFT.
}
\end{figure}

From fits to Eq.~(\ref{eq:ppole}) we determine the form factor radii $r_1$
and $r_2$ as well as the anomalous magnetic moment $\kappa$.
The quark mass dependence of these quantities has been calculated using the
SSE formalism \cite{ff-sse}. For $r_1$ the parameters are known and we
therefore restrict ourselves to a comparison of the SSE result and the
lattice data (see Fig.~\ref{fig:r1v}). While for $\mps \gtrsim 300\,\mbox{MeV}$
the lattice results are significantly smaller than the phenomenological value,
towards smaller quark masses we observe an increase of the radius.
This is consistent with predictions from ChEFT. For $r_2$ and $\kappa$
we find a similar behaviour. Since there are no phenomenological
estimates for all parameters of the SSE expressions we perform a
combined fit.  The results are plotted in \ref{fig:r2v} and \ref{fig:kv}.

\vspace*{-1mm}
\section{Summary and outlook}
\vspace*{-1mm}

We have presented an update of QCDSF results on the lowest moments of
unpolarized, polarized and tensor PDFs as well as the electromagnetic
form factors. Some of our results at light quark masses with
$\mps \lesssim 300\,\mbox{MeV}$ confirm the expectations from ChEFT that
light quark mass effects are significant. However, this possibly does not
explain all of the observed discrepancies from phenomenological values.

\section*{Acknowledgements}

The numerical calculations have been performed
on the APEmille and apeNEXT systems at NIC/DESY (Zeuthen),
the BlueGene/P at NIC/JSC (J\"ulich),
the BlueGene/L at EPCC (Edinburgh),
the Dell PC-cluster at DESY (Zeuthen),
the QPACE systems \cite{qpace} of the SFB TR-55,
the SGI Altix and ICE systems at LRZ (Munich) and HLRN (Berlin/Hannover).
This work was supported in part
by the DFG (SFB TR-55) and
by the European Union (grants 238353, ITN STRONGnet and 227431, HadronPhysics2,
and 256594).


\end{document}